\documentclass[review,endfloats,12pt]{elsarticle}

\usepackage{amssymb}
\usepackage{siunitx}

\newlength\figurewidth
\makeatletter
\renewcommand\listoffigures{\@starttoc{}}
\renewcommand\listoftables{\@starttoc{}}
\makeatother

\journal{Nucl. Instrum. Meth. B}

\date{accepted for publication, 3.6.2020}

\begin{document}

\begin{frontmatter}

\title{Temporally resolved LEIS measurements of Cr segregation after preferential sputtering of WCrY alloy}

\author[fzj]{H. R. Koslowski\corref{cor}}
\ead{h.r.koslowski@fz-juelich.de}
\author[fzj,gent]{J. Schmitz}
\ead{jan.schmitz@fz-juelich.de}
\author[fzj]{Ch. Linsmeier}
\ead{ch.linsmeier@fz-juelich.de}
\address[fzj]{Forschungszentrum J\"ulich GmbH, Institut f\"ur Energie- und Klimaforschung -- Plasmaphysik, 52425 J\"ulich, Germany}
\address[gent]{Department of Applied Physics, Ghent University, 9000 Ghent, Belgium}
\cortext[cor]{Corresponding author}

\begin{abstract}
The dynamic behaviour of thermally driven segregation of Cr to the surface of WCrY smart alloy is studied with low energy ion scattering (LEIS).
Sputtering the WCrY sample with \SI{500}{eV} D$_2^+$ ions at room temperature results in preferential removal of the lighter alloy constituents and causes an almost pure W surface layer.
At elevated temperatures above \SI{700}{K} the segregation of Cr atoms towards the surface sets in and prevents the formation of a pure W layer.
The simultaneous heating and sputtering of the sample leads to a surface state which reflects the balance between sputter removal and segregation flux, and deviates from the equilibrium due to thermally driven segregation.
Stopping the sputter ion beam allows the system to relax and develop toward the segregation equilibrium.
The time constants for the temporal changes of W and Cr surface coverage are obtained from a series of LEIS measurements.
The segregation enthalpy is determined from the time constants obtained for various sample temperatures.
\end{abstract}


\end{frontmatter}

\section{Introduction}

The investigation of candidate materials for the first wall of a future fusion power plant is an ongoing topic in nuclear fusion research.
One of the most suitable materials is tungsten due to its high melting point and small sputter erosion under particle bombardment \cite{bol02}\cite{neu16}.
However, the tungsten wall cladding in a fusion environment is exposed to a high neutron flux.
Stable tungsten isotopes can undergo transmutation forming e.g. rhenium and radioactive isotopes of tungsten.
For example, the stable tungsten isotopes $^{182}$W and $^{186}$W have rather large cross sections for the (n,2n) reaction \cite{qai75} and abundances of the resultant isotopes will increase with operational time of the power plant.
In case of a loss-of-coolant accident accompanied by air or water ingress the temperature of the first wall can rise up to \SIrange{900}{1200}{\degreeCelsius} \cite{mai05}.
Under these conditions, tungsten and its main transmutation products rhenium and osmium form oxides which are volatile above \SI{700}{\degreeCelsius}, and are released into the vacuum vessel.

A possible route to prevent the release of radioactive material is the application of self-passivating tungsten alloys \cite{koc11}.
The self-passivating alloy contains chromium (and other constituents).
Under accidental conditions the chromium forms a stable oxide layer on the surface which prevents the further oxidation and release of tungsten, whereas preferential sputtering of the lighter alloy constituents by bombarding plasma particles causes the enrichment of tungsten on the surface which results in little erosion under normal operation.

A self-passivating WCrY alloy has been recently developed and the oxidation resistance has been verified in laboratory experiments \cite{lit17}.
Furthermore, the development of a tungsten rich surface has been confirmed by plasma exposure in the linear plasma device PSI-2 and subsequent secondary ion mass spectrometry analysis of the surface composition \cite{sch18}.

A recent investigation of WCrY alloy with LEIS has been focused on the thermal stability and the preferential sputtering \cite{kos20}.
The experiment confirmed that ion beam bombardment with \SI{250}{eV} deuterons at room temperature produces an almost pure tungsten surface due to preferential sputtering.
However, when the sample temperature of an unexposed sample is increased above \SI{700}{K} the segregation of chromium towards the surface sets in, and at a temperature of \SI{1000}{K} no tungsten is found in the topmost surface layer.

In this work we will continue the previous study and investigate the dynamic behaviour of the temperature driven segregation.
For that purpose, sputter erosion and sample heating are applied simultaneously in order to produce a non-equilibrium state of the sample surface.
It neither shows the full tungsten enrichment due to preferential sputtering, nor does the surface show the segregation equilibrium depending on the sample temperature.
Stopping the ion bombardment allows the surface to relax under action of the segregating particle flux.
A continuous LEIS measurement of the individual alloy constituents yields time-resolved surface concentrations.

\section{Experiment}

The LEIS apparatus has been described in detail in \cite{kos18}.
Singly charged ions (typically He$^+$ or Ne$^+$) are produced in a Bayard-Alpert type ion source, extracted with a weak electric field, and accelerated to an energy of typically \SI{1}{keV}.
The ion beam is focused and steered onto the sample by two sets of einzel lenses and deflection plates.
The sample holder of type Prevac PTS 1000 RES/C-K is mounted at a four axis manipulator (VG Scienta).
The sample can be rotated to allow for incident angles in the range \SIrange{0}{90}{\degree} and heated up to temperatures in excess of \SI{1200}{K}.
Reflected ions are analysed with an electrostatic \SI{90}{\degree} spherical sector analyser and detected with a channeltron electron multiplier in single pulse counting mode.
The counting and the control of the potentials of the electrostatic analyser are done with a National Instruments PXI-8106 data acquisition board which is controlled by LabVIEW software.
A Perkin--Elmer 04--161 sputter ion gun serves for sample cleaning and sputter erosion.
The base pressure in the apparatus is less than $\SI{5e-10}{mbar}$.

\begin{figure}[t]
\centering\includegraphics[width=\figurewidth]{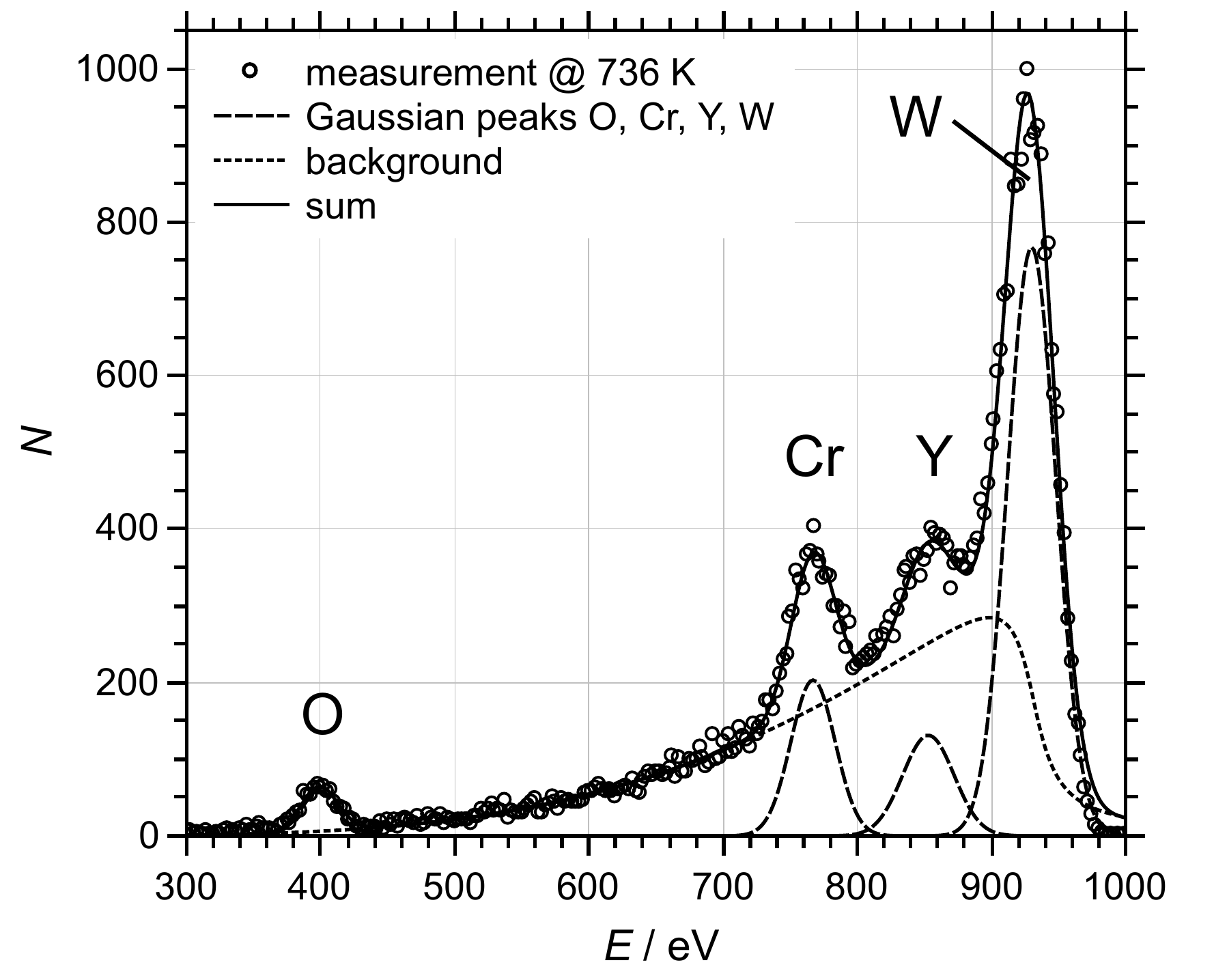}
\caption{LEIS energy spectrum of \SI{1}{keV} He$^{+}$ scattered on WCrY measured at \SI{736}{K}.
The peaks for O (\SI{400}{eV}), Cr (\SI{760}{eV}), Y (\SI{845}{eV}), and W (\SI{925}{eV}) are fitted with Gaussians (dashed lines), and the re-ionisation background from W is fitted with a modified semi-empirical formula (dotted line) \cite{nel86}.
The sum of all contributions (black line) shows good agreement with the measurement (open circles).}
\label{fig:spec}
\end{figure}

A WCrY ingot is produced by a field-assisted sintering technology \cite{lit17}.
The alloy has a composition of \SI{88.0}{wt\%} W, \SI{11.4}{wt\%} Cr, and \SI{0.6}{wt\%} Y.
The corresponding atomic fractions are \SI{67.9}{at\%} W, \SI{31.1}{at\%} Cr, and \SI{1.0}{at\%} Y.
The WCrY sample for the LEIS measurements is cut by wire erosion and polished to a mirror finish.
More details on the sample preparation can be found in \cite{kos20}.

Prior to the LEIS measurements the sample surface is cleaned by sputtering with \SI{500}{eV} Kr$^+$ ions with an integrated charge of several \si{mC} on an area of about \SI{0.25}{cm^2}.
The sample cleaning causes already an unavoidable W surface enrichment because the elemental sputter yields for Cr and W are \num{1.0} and \num{0.6}, respectively \cite{yam96}.
However, this effect is rather small compared to the preferential sputtering caused by D$_2^+$ ions.

It cannot be excluded that the cleaning by Kr ion sputtering as well as the preferential sputtering by D ions results in a modification of the surface morphology of the WCrY sample which has an influence on the ion scattering signals \cite{jan04}.
Since the main result of this work, e.g. the temporal evolution of the surface composition (shown in figure \ref{fig:seg-vs-t}), has been measured after sputter cleaning and preferential erosion on the same surface, any influence from morphology and surface roughness might impact all measurements in the same way.
Furthermore, surface roughness will mainly influence the ion scattering signal intensity and is unlikely to change the ratio of scattering peaks arising from different elements in the alloy.

Figure \ref{fig:spec} shows a typical measurement of the energy distribution of He$^+$ ions which are reflected from the WCrY sample surface.
The incident ion energy is \SI{1}{keV}.
The probing ion beam strikes the surface under an angle of \SI{20}{\degree} with respect to the surface normal direction.
The reflected ions are measured under a scattering angle of \SI{140}{\degree}.
The sample is kept at a temperature of \SI{736}{K} during the measurement.
The spectrum (open circles) shows four distinct peaks which are attributed to the scattering of the He ions at O, Cr, Y, and W atoms.
Besides the peaks there is a pronounced background which arises due to the re-ionisation of neutralised He atoms which scatter at a W atom.
This background is well described by a modified semi-empirical formula given originally by Nelson \cite{nel86}.
\begin{equation}
    \begin{array}{rcl}
    f_{\rm bg}(E) & = & B (\pi/2 - \arctan{(S (E-E_0)/w))} \\
                  &   & \times \exp(-K \sqrt{E^{\prime}})\\
    E^{\prime} & = & \Bigg\{
        \begin{array}{ll}
            E   & {\rm if\ } E \leq E_0 \\
            E_0 & {\rm if\ } E > E_0
        \end{array}
    \end{array}
    \label{eq:bg}
\end{equation}
Equation \ref{eq:bg} shows the used formula for fitting the re-ionisation background.
Here, $E_0$ is the energy of the W peak, $w$ is the full width at half maximum, and $B$ and $K$ are free parameters.
In order to obtain a better match between measurement and fit, the slope of the inverse tangent function at the peak position is made variable by another fitting parameter $S$.
Furthermore, the argument of the $\exp$ function is kept constant above the peak energy to facilitate the decay of the background towards energies above the peak centre.
The peaks originating from scattering at the various surface atom types are fitted with Gaussian profiles.
The sum (full line) of the four peaks (dashed lines) and the re-ionisation background (dotted line) results in a good approximation of the measured spectrum.

The relative surface concentration of a species $i$ in the first atomic layer is calculated as
\begin{equation}
    c_{i} = \frac{A_{i}}{\Sigma A_{j}},
    \label{eq:conc}
\end{equation}
where the sum is taken over all species $j$ detected on the surface. The quantities $A_{i,j}$ denote the areas of the fitted peaks of the respective species.
Equation \ref{eq:conc} is strictly valid only when the elemental sensitivity factors of the various elements on the surface are equal which is normally not the case.
However, the total number of counts in the LEIS spectra shown below in figure \ref{fig:seg-vs-t} exhibit a variation of $\pm 10\%$ around the mean value despite much larger changes of the various surface concentrations.
The moderate change in total counts indicates that elemental sensitivity factors of the varying surface constituents are not much different and supports the validity of equation \ref{eq:conc}.

The preferential sputter erosion of the surface is realised with \SI{500}{eV} D$_2^+$ ions.
For \SI{250}{eV} deuterons the elemental sputter yields for W and Cr are \num{3e-4} and \num{0.04}, respectively \cite{sug16}.
Simultaneous heating of the sample results in an equilibrium state which depends on the balance between the flux of eroding particles and the temperature dependent flux of segregating Cr atoms towards the surface.

\begin{figure}[t]
\centering\includegraphics[width=\figurewidth]{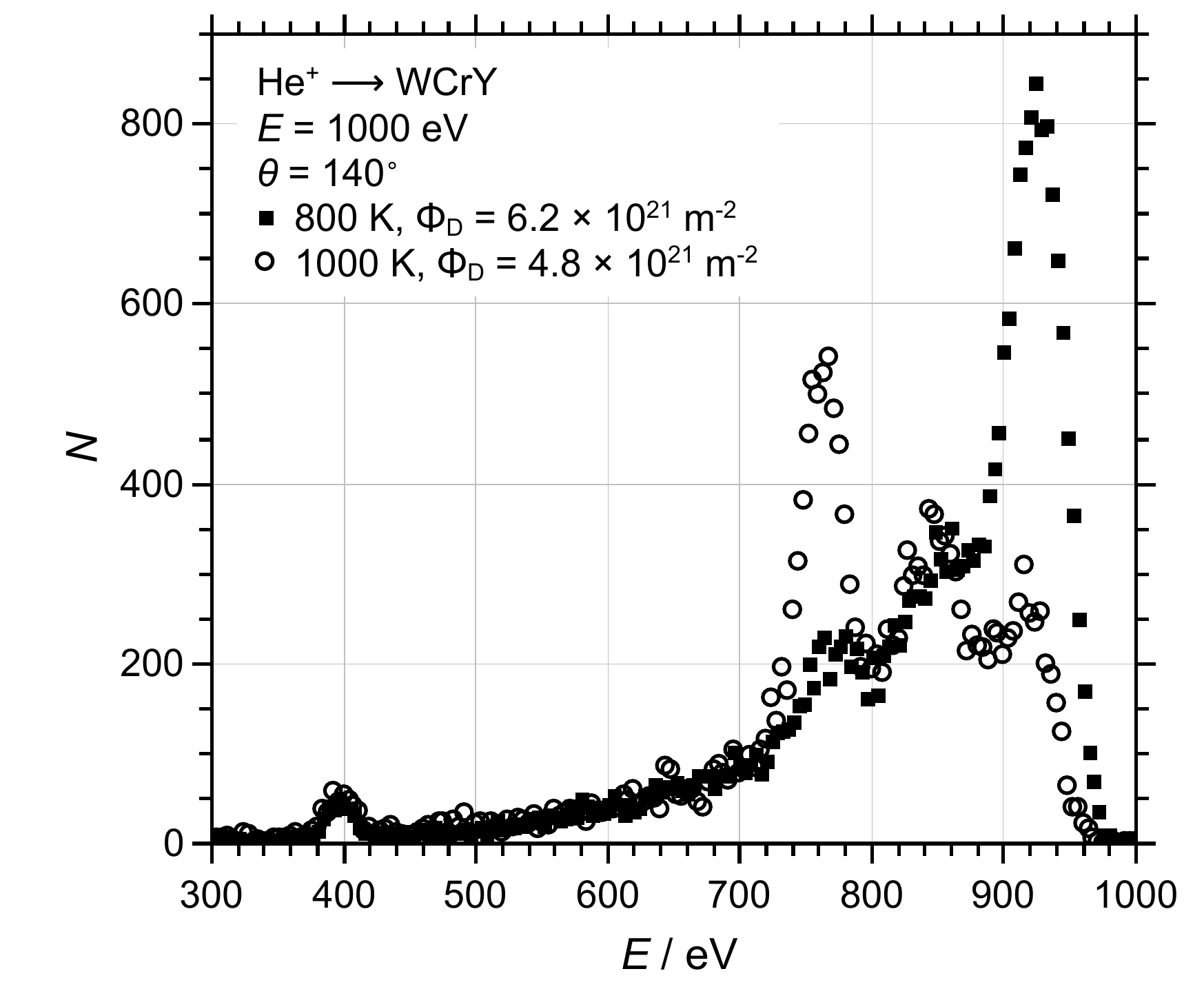}
\caption{LEIS spectra measured after preferential sputtering with \SI{250}{eV} deuterons at elevated sample temperatures.
At \SI{800}{K} (black squares) the surface layer consists mainly of W, but the Cr peak (at \SI{760}{eV}) is still present.
At \SI{1000}{K} (open circles) the Cr segregation is rather strong and prevents the development of a pure W surface layer.}
\label{fig:sput-seg}
\end{figure}

Figure \ref{fig:sput-seg} shows LEIS spectra measured within a few minutes directly after stopping the preferential sputtering cycle at two different temperatures.
In both cases the preceding sputtering cycles are done with a D flux of \SI{1e18}{m^{-2}s^{-1}} and sufficient fluences in order to achieve an almost complete removal of the lighter alloy constituents if the sample would have been exposed at room temperature \cite{kos20}.
Since the sample is kept at elevated temperature the obtained surface compositions are different.
At a temperature of \SI{800}{K} (black squares) the W peak at \SI{925}{eV} is still dominant.
However, the peaks from scattering on Cr (\SI{760}{eV}) and Y (\SI{845}{eV}) are already present, proving that temperature driven segregation towards the surface prevents the formation of a pure W surface layer.
The second measurement shown in figure \ref{fig:sput-seg} is taken at an higher temperature of \SI{1000}{K} (open circles).
Under these conditions the pure Cr segregation (without concurrent sputtering) would lead to a complete coverage of the topmost surface layer, suppressing scattering of the incident ions on W atoms \cite{kos20}.
However, the data still shows a pronounced peak at \SI{925}{eV} due to single scattering on W.
Both measurements indicate that under conditions with sputter erosion at elevated temperatures a state in a dynamic equilibrium develops.
\begin{table}[h]
\begin{tabular}[h]{ r r r r r r r }
\hline
$T / {\rm K}$ & $\Phi_{\rm D} / \si{m^{-2}}$ & $c_{\rm O}$ & $c_{\rm Cr}$ & $c_{\rm Y}$ & $c_{\rm W}$ & Ref. \\
\hline
 796 & 0            & 0.048 & 0.244 & 0.106 & 0.602 & fig. 3 in \cite{kos20} \\
 800 & \num{6.2E21} & 0.030 & 0.082 & 0.115 & 0.773 & this work \\
 981 & 0            & 0.016 & 0.744 & 0.230 & 0.011 & fig. 3 in \cite{kos20} \\
1000 & \num{4.8E21} & 0.046 & 0.530 & 0.277 & 0.147 & this work \\
\hline
\end{tabular}
\caption{\label{tab:conc}
Surface concentrations of the various constituents (Cr, Y, W) and O impurity at temperatures of {\SI{800}{K}} and {\SI{1000}{K}}. The compositions of an only heated sample (lines 1 and 3) is compared to the heated under D sputtering sample (lines 2 and 4).}
\end{table}
The obtained surface composition is between the segregated state in the absence of any sputter erosion and the state with W enrichment due to preferential sputtering which builds up at room temperature.
This behaviour is documented in table \ref{tab:conc} where the surface compositions of the sample at temperatures of \SI{800}{K} and \SI{1000}{K} are compared for the cases with heating only (which gives the equilibrium concentrations) and simultaneous heating combined with \SI{250}{eV} deuteron sputtering.

For the rest of this paper we neglect the apparent segregation of Y which occurs always in parallel to the Cr segregation.
Y is contained in the WCrY alloy mainly in the form of yttria nanoparticles which are located at the grain boundaries of larger W-Cr grains and do not form part of the solid solute \cite{lit18}.
Although Y seemingly shows segregation similar to Cr, the underlying mechanism is likely different.

\section{Results and Discussion}

\begin{figure}[t]
\centering\includegraphics[width=\figurewidth]{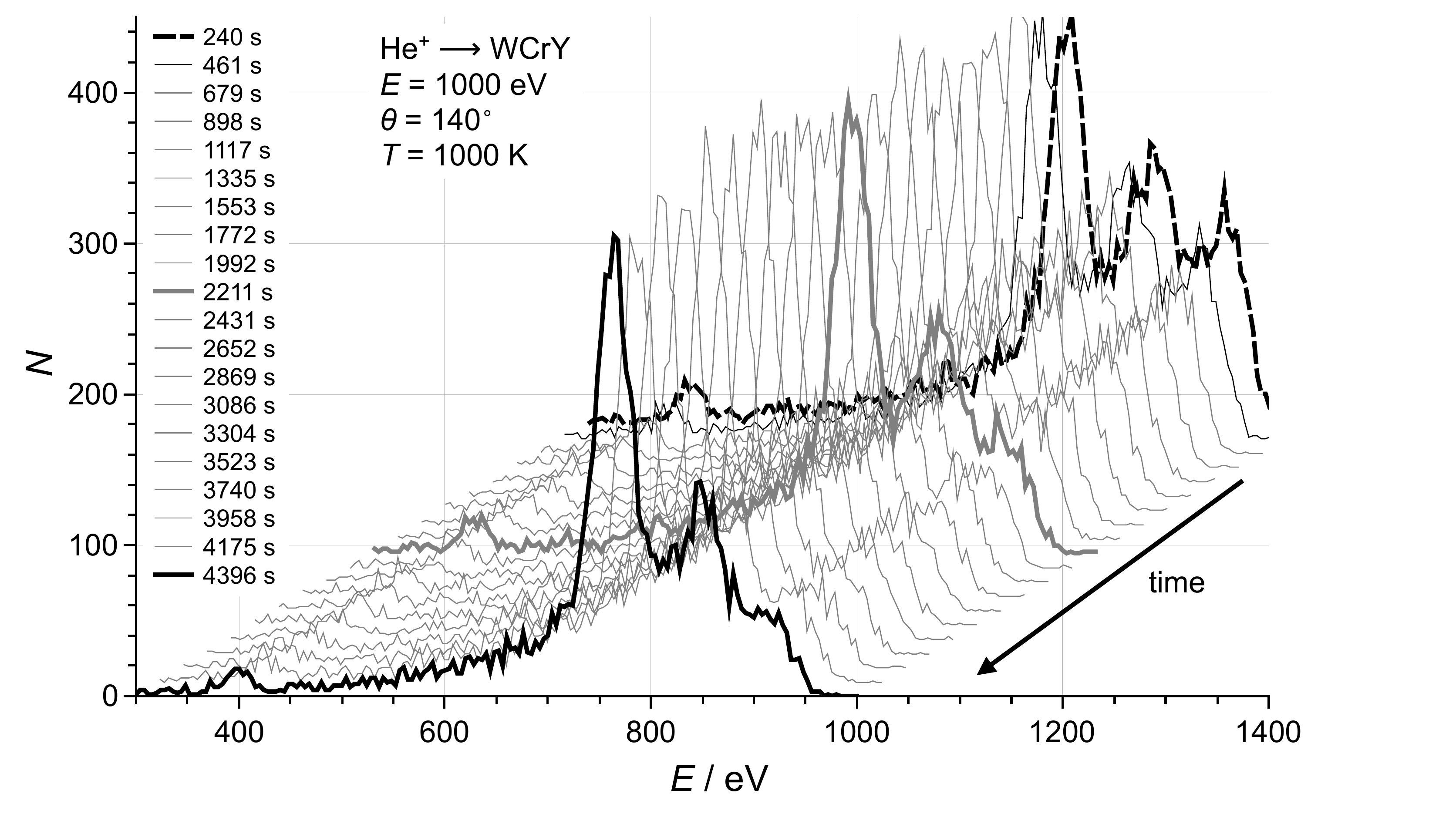}
\caption{Sequence of LEIS spectra measured after preferential sputtering of a WCrY sample by \SI{250}{eV} deuterons with a fluence of $\Phi_{\rm D} = \SI{4.8E21}{m^{-2}}$.
The sample is kept at a temperature of \SI{1000}{K}.
The spectra are recorded in quick succession after stopping the deuteron irradiation.
The first measured spectrum is shown at the back of the waterfall plot, the last measured spectrum is at the front.}
\label{fig:seg-vs-t}
\end{figure}

In the following the temporal evolution of the Cr surface coverage due to thermally activated segregation will be investigated.
For that purpose the WCrY sample is heated up to a temperature of \SI{1000}{K} and kept at this temperature with simultaneous sputter erosion by \SI{250}{eV} deuterons by irradiation with \SI{500}{eV} D$_2^+$ ions.
The sputter cycle lasts for \SI{4300}{s} allowing the sample to reach the dynamic equilibrium.
Immediately after stopping the sputter ion beam a series of LEIS spectra is measured.
Twenty spectra are measured in total and the results are summarised in figure \ref{fig:seg-vs-t}.
The measurements are taken continuously.
The integration time per spectrum is reduced to \SI{3.5}{min}.
Although the number of recorded counts per energy channel is reduced, the spectra contain sufficient statistical relevance to allow stable peak fitting.
All spectra are shown in a waterfall plot.
The first spectrum is at the back of the plot and time proceeds towards the front.
The time intervals between the end of the sputter cycle and any of the measurements is indicated in the left part of the figure.
The spectrum measured first (dashed thick black line, it is the same spectrum as shown in figure \ref{fig:sput-seg} with open circles) has a pronounced W peak at the right side.
This peak gradually shrinks when time proceeds, as can be seen by the intermediate spectrum taken at \SI{2211}{s} (thick grey line) and the last measured spectrum in the front (full thick black line).
Corresponding to the decay of the W surface concentration the Cr peak increases continuously.
It has to be noted that the \SI{1}{keV} He$^{+}$ LEIS probing beam, although the beam current of approximately \SI{3}{nA} is very small, results in some unavoidable preferential sputtering of Cr.
The Cr sputter yield calculated according to \cite{eck07} and using the measured beam width amounts to $\SI{4.2e14}{m^{-2}s^{-1}} \times c_{\rm Cr}$.
In order to estimate the influence on the temporal evolution of the Cr surface concentration this value can be compared to the sputter yield of the \SI{250}{eV} deuterons which amounts to $\SI{4e16}{m^{-2}s^{-1}} \times c_{Cr}$ and is about 100 times larger.
In the dynamic equilibrium between sputtering and segregation this yield is balanced by the segregation flux of Cr atoms toward the surface.
From these numbers one can conclude that the effect of sputtering by the probing beam ions remains rather small.

\begin{figure}[t]
\centering\includegraphics[width=\figurewidth]{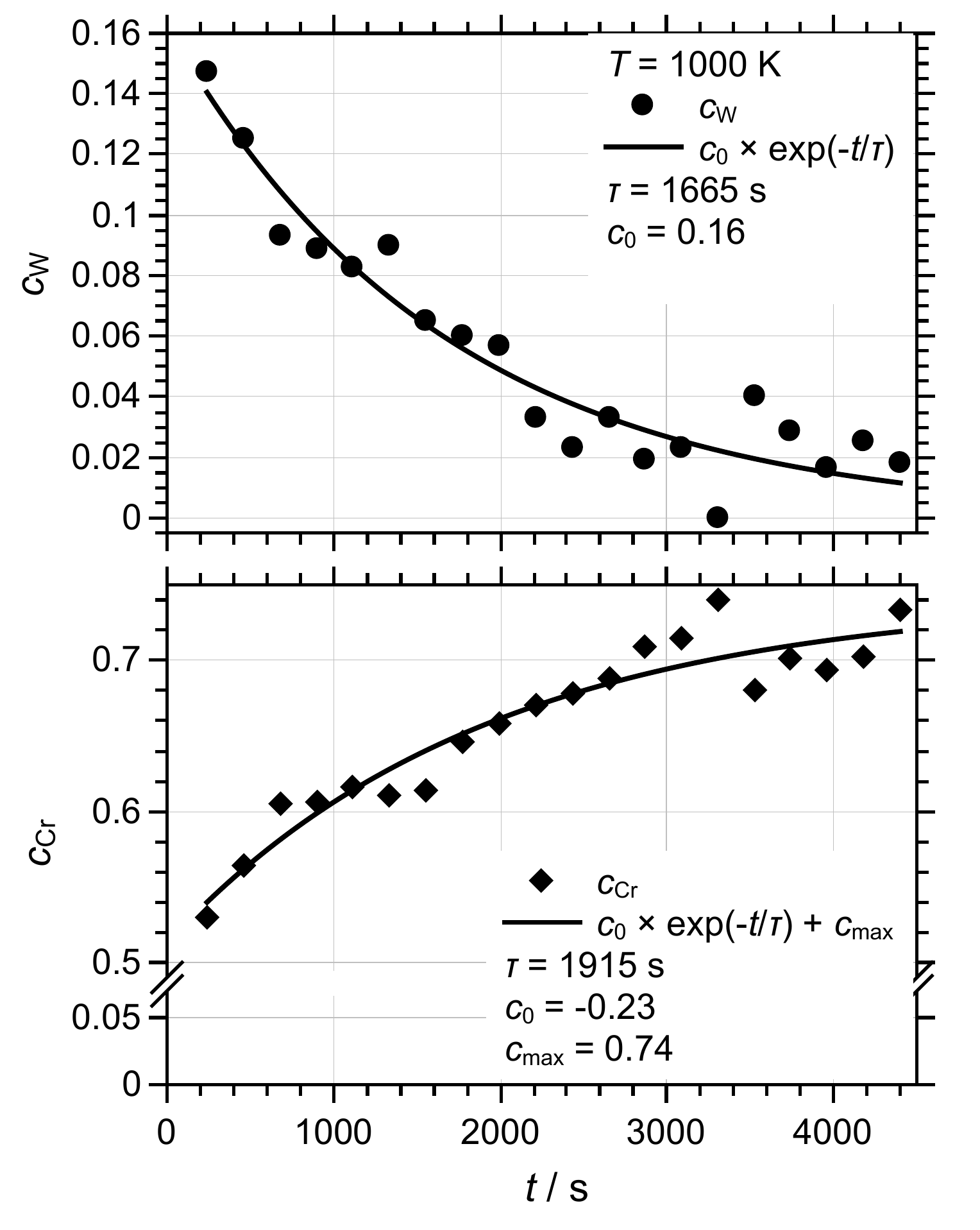}
\caption{Relative surface concentrations of W, $c_{\rm W}$, (top) and Cr, $c_{\rm Cr}$, (bottom) plotted versus time.
The values are derived according to equation \ref{eq:conc} by peak fitting the spectra shown in figure \ref{fig:seg-vs-t}.}
\label{fig:w-cr-conc}
\end{figure}

All measured spectra shown in figure \ref{fig:seg-vs-t} are fitted by a sum of four Gaussian peaks representing the scattering from W, Cr, Y and O, and the modified semi-empirical expression for the re-ionisation background as depicted in figure \ref{fig:spec}.
Relative surface concentrations of the various constituents are calculated according to equation \ref{eq:conc}.
The results for W and Cr are plotted in figure \ref{fig:w-cr-conc}.
The W surface concentrations (top, black dots) exhibit a temporal decay due to the segregation of Cr to the surface.
The data show a rather large scatter caused by the relatively short integration time per spectrum which has been chosen in order to allow for a sufficient temporal resolution of the surface changes.
The line shows an exponential decay
\begin{equation}
    c_{\rm W}(t) = c_0 \times \exp(-t/\tau_{\rm W})
    \label{eq:wvst}
\end{equation}
fitted to the data, respecting the asymptotic boundary condition that in the fully segregated state the W concentration in the topmost surface layer is below the detection limit \cite{kos20}.
The agreement between measured surface concentrations and the fitted curve is reasonably well taking into account the somewhat noisy data due to the short integration time.
Figure \ref{fig:w-cr-conc} shows at the bottom the same plot for the relative Cr surface concentration.
The Cr concentration increases over time and shows saturation at a value less than \num{1} due to the concurrent (apparent) segregation of Y and the small O peak which is always present in the measured spectra.
Again, the line shows a fit to the data which is substantiated by the assumption that the probability of a Cr atom to occupy a surface position is proportional to the difference between the maximum (asymptotic) surface fraction $c_{max}$ and the actual surface concentration $c_{\rm Cr}$
\begin{equation}
    \frac{{\rm d}c_{\rm Cr}(t)}{{\rm d}t} \propto (c_{\rm max} - c_{\rm Cr}(t)),
    \label{eq:diff}
\end{equation}
motivated by the ansatz made in \cite{swa81}.
The solution of the differential equation \ref{eq:diff} is
\begin{equation}
    c_{\rm Cr}(t) = c_0 \times \exp(-t/\tau_{\rm Cr}) + c_{\rm max},
    \label{eq:crvst}
\end{equation}
whose fit to the data yields good agreement.
Note that the coefficient $c_0$ is negative because the solution describes a monotonously increasing function which asymptotically approaches the value $c_{\rm max}$.
The time constant for Cr segregation to the surface and filling up surface positions amounts to \SI{1915}{s}.

The time constants for the W decrease and the Cr increase are long and it takes several hours to recover from the perturbed surface state due to the preferential sputtering.
However, both determined time constants are slightly different.
The deviation might be due to the concurrent precipitation of Y or to some extent caused by the rather noisy data.

A repetition of the same experiment at a sample temperature of \SI{800}{K} yields a much larger time constant of \SI{7074}{s} for the Cr segregation.
Table \ref{tab:tau} summarises the obtained time constants $\tau_{\rm Cr}$ and $\tau_{\rm W}$ for Cr increase and W decrease on the surface.
The rather large error margin of the $\SI{800}{K}$ measurement might be caused by a too short measurement time of the order of the time constant, which complicates the numerical fit of exponential functions \ref{eq:wvst} and \ref{eq:crvst} to the data.

\begin{table}[h]
\begin{tabular}[h]{ r r r r r }
\hline
$T$ / K & $\tau_{\rm Cr}$ / s & $\Delta \tau_{\rm Cr}$ / s & $\tau_{\rm W}$ / s & $\Delta \tau_{\rm W}$ / s \\
\hline
800 & 7074 & ($\pm$ 5526) & 4252 & ($\pm$ 2264) \\
1000 & 1915 & ($\pm$  547) & 1665 & ($\pm$ 141) \\
\hline
\end{tabular}
\caption{\label{tab:tau}
Time constants of Cr segregation, $\tau_{\rm Cr}$, and corresponding time constants for W decrease, $\tau_{\rm W}$ for sample temperatures of \SI{800}{K} and \SI{1000}{K}.
The errors in the third and fifth column are determined in the usually way from the diagonal elements of the covariance matrix.}
\end{table}

The solution (equation \ref{eq:crvst}) for the temporal evolution of the surface concentration of the segregating species, Cr, can be rewritten as
\begin{equation}
    c_{\rm vacant} = c_{\rm max} - c_{\rm Cr}(t) = -c_0 \times \exp(-t/\tau_{\rm Cr}),
    \label{eq:1storder}
\end{equation}
where $c_{\rm vacant}$ is the concentration of vacant surface positions which can be occupied by segregating Cr atoms.
This equation has formally the same structure as a first order chemical reaction equation where the reaction rate constant $k$ equals the inverse of the time constant $1/\tau_{\rm Cr}$.
The corresponding Arrhenius plot of $k = 1/\tau_{\rm Cr} = f(1/T)$ yields the activation energy which can be equated with the segregation enthalpy $\Delta H_{\rm seg}$.
Using the values from table \ref{tab:tau} and calculating the slope of the straight line defined by both points gives a segregation enthalpy $\Delta H_{\rm seg} = 0.45 (\pm 0.37) \si{eV}$.
This value is compatible with the value of \SI{0.7}{eV} which has been determined from the Langmuir-McLean isotherm in a previous publication \cite{kos20}.

\section{Summary and Conclusion}

Simultaneous heating and preferential sputtering by D bombardment of a WCrY alloy sample results in a non-equilibrium state which is somewhere in between the states where either Cr is fully segregated to the surface, depending on the temperature, or a fully W covered surface is obtained at room temperature.
Stopping the sputter erosion lets the sample relax to the segregation equilibrium.

The analysis of a series of LEIS measurements documents the temporal evolution of the surface concentrations of the various alloy constituents.
The time constant for Cr segregation at \SI{1000}{K} is \SI{1915}{s^{-1}}, i.e. the relaxation after perturbation by preferential erosion occurs on a timescale of several hours.
The analysis of the segregation time constants allows to determine the segregation enthalpy.

The experimental results show that the segregation of Cr to the surface of the WCrY sample induces a strong change in the surface composition.
The ion beam laboratory experiment monitors such changes in-situ.
With respect to often applied plasma loading with subsequent sample analysis one has to be aware that the post-mortem measured surface composition most likely reflects not the surface composition under plasma loading at elevated temperature but rather after establishing an equilibrium surface concentration after termination of the plasma flux to the surface.
For example the quantification of erosion should take into account that the segregating species is stronger eroded, especially in cases where it is a lighter alloy constituent with enhanced sputter yield compared to alloy constituents with higher masses.

\section*{Acknowledgements}

We thank Mr. Albert Hiller and Mr. Roland B\"ar for technical assistance and maintenance of the LEIS apparatus, and Mrs. Beatrix G\"oths for polishing the WCrY sample.

This work has been carried out within the framework of the EUROfusion Consortium and has received funding from the Euratom research and training programme 2014-2018 and 2019-2020 under grant agreement No 633053. The views and opinions expressed herein do not necessarily reflect those of the European Commission.

\end{document}